\documentclass[aps,preprint,showpacs,preprintnumbers,amsmath,amssymb]{revtex4}
\usepackage[dvips]{graphicx}
\usepackage{float}

\begin{document}

\title{Annihilation of the scalar pair into a photon on de Sitter space-time}
\author{Mihaela-Andreea B\u aloi}

\email{mihaela.baloi88@e-uvt.ro}
 \affiliation{Faculty of Physics, West University of Timi\c soara,  \\V. P\^ arvan
 Avenue 4, RO-300223 Timi\c soara,  Romania}

\begin{abstract}
The annihilation of massive scalar particles in one photon on de Sitter expanding universe is studied, using perturbation theory. The amplitude and probability corresponding to this process is computed using the exact solutions of the Klein-Gordon and Maxwell equations on de Sitter geometry. Our results show that the expression of the total probability of photon emission is a function dependent on the ratio $mass/expansion\, factor$. We perform a graphical study of the total probability in terms of the parameter $mass/expansion factor$, showing that this effect is significant only in strong gravitational fields. We also obtain that the total probability for this process vanishes in the Minkowski limit.
\end{abstract}


\maketitle

\section{Introduction}
The theory of quantum fields on curved space-times has lead to many important results.
Starting with the paper of Schr\"odinger from 1939 related to the production of scalar particles as a result of the expansion of the background, many authors have turned their attention to this subject, using different approaches \cite{B1}\cite{B3}\cite{B24}\cite{B7}-\cite{B10}. One of the most studied subjects is related to the production of scalar particles on Friedmann-Robertson-Walker (FRW) space-times \cite{B1}\cite{B26}\cite{B10}. The studies of Parker based on minimally coupled field equations, showed that particles are produced in pairs and this process is significant only in the strong gravitational fields of the early Universe \cite{B1}-\cite{B3}. A similar result was obtained in Ref.\cite{B10} where two different mechanisms of particle creation were compared: the production of electromagnetically interacting particles and free particle creation i.e nonconformally coupled scalar field with gravity. This study led to the conclusion that the first type of particle creation, which is based on perturbation theory, is dominant in the early Universe \cite{B10}.
We also mention that a detailed study of fermion pair production from a photon and the time reversed process in the FRW metric with conformal flat line element can be found in Ref.\cite{B25}. In the more recent papers which study the production of particles on de Sitter space-time, based on the $S$ matrix approach \cite{B7}\cite{B9}, the results also prove that the effect of particle production is important only for large expansion factor corresponding to the early Universe.

This letter is focused in the study of the scalar pair annihilation into a photon $\varphi+\varphi^{*}\rightarrow\,\gamma$ on de Sitter geometry. The time reversed process is obtained when the scalar pair is generated by a photon. We have denoted by $\varphi$  the scalar particle, $\varphi^{*}$ the scalar antiparticle and $\gamma$ the photon. It is known that, the QED formalism from Minkowski space can be adapted for the study of QED processes in curved spacetimes \cite{B9}\cite{B11}. Then the definition of the transition amplitude in de Sitter geometry can be established \cite{B24}-\cite{B7}\cite{B9}. In our study, the $in$ and $out$ fields are  exact solutions of the Klein-Gordon and Maxwell equations in de Sitter space, written in the momentum-helicity basis.

In section II we present the main steps for computing the expression of the probability density for scalar pair annihilation in one photon. Section III is dedicated to the study of the total probability of the created photons. This is one of the few studies in which the total probability can be calculated exactly. Further we graphically analyse the behaviour of the total probability of photon emission as function of the parameter $m/\omega$, and given values for the momenta of the scalar particles. In section IV we present the conclusions.

\section{The transition amplitude}
The expression of the line element which describes the de Sitter universe is \cite{B19}:
\begin{equation}
ds^{2}=dt^{2}-e^{2\omega t}d\vec{x}\,^{2}=\frac{1}{(\omega t_{c})^{2}}(dt_{c}^{2}-d\vec{x}^{2}),
\end{equation}
where $\omega>0$ is the expansion factor and the conformal time is $t_{c}=-\frac{e^{-\omega t}}{\omega}$, $t_{c}\in(-\infty, 0)$. In order to study the annihilation of the scalar pair into a photon in de Sitter geometry, we will use the solution of the Klein-Gordon equation from Ref.\cite{B5}. Therefore the positive frequency solution of the Klein-Gordon equation, in momentum basis is \cite{B5}:
\begin{equation}
f_{\vec{p}}\,(x)=\frac{1}{2}\sqrt{\frac{\pi}{\omega}}\frac{e^{-3\omega t/2}}{(2\pi)^{3/2}}e^{-\pi k/2}H^{(1)}_{ik}\left(\frac{p}{\omega}e^{-\omega t}\right)e^{i\vec{p}\,\vec{x}}, \label{a1}
\end{equation}
where $H^{(1)}_{ik}$ is the Hankel function of the first kind and $k=\sqrt{x^2-\frac{9}{4}}$, where $x=\frac{m}{\omega}>\frac{3}{2}$.\\
The theory of electromagnetic field in de Sitter metric was studied in Refs.\cite{B4}-\cite{B15}. In Ref.\cite{B4} is constructed the theory of the free Maxwell field on de Sitter space-time in the conformal chart $\{t_{c},\vec{x}\}$ and Coulomb gauge i.e. $A_{0}=0$, $(\sqrt{-g}A^{i})_{;i}=0$. Using the conformal invariance, the positive frequency solution of the free electromagnetic field in momentum-helicity basis can be easily obtained \cite{B4}:
\begin{equation}
w^{i}_{\vec{q},\lambda}(x)=\frac{1}{(2\pi)^{3/2}}\frac{1}{\sqrt{2q}}e^{-iqt_{c}+i\vec{q}\,\vec{x}}\,\varepsilon_{\lambda}^{i}(\vec{q}\,)e^{-2\omega t}, \label{a2}
\end{equation}
where $\vec{\varepsilon}_{\lambda}(\vec{q}\,)$ is the polarization vector, which is orthogonal on the momentum of the photon $\vec{q}\cdot\vec{\varepsilon}_{\lambda}(\vec{q}\,)=0$,  and satisfies the following relations \cite{B4}:
\begin{eqnarray}
\vec{\varepsilon}_{\lambda}(\vec{q}\,)\vec{\varepsilon}\,^{*}_{\lambda\,'}(\vec{q}\,)&=&\delta_{\lambda\lambda\,'}\nonumber\\
\sum_{\lambda}\vec{\varepsilon}_{\lambda}(\vec{q}\,)_{i}\vec{\varepsilon}\,^{*}_{\lambda}(\vec{q}\,)_{j}&=&\delta_{ij}-\frac{q^{i}q^{j}}{q^{2}}.\label{b2}
\end{eqnarray}

Further we are interested in studying the annihilation of the scalar pair into a photon using the perturbative methods adapted to de Sitter geometry. It is worth mentioning that this process is forbidden in the scalar QED from Minkowski space, because of the simultaneous conservation of energy and momentum. The transition amplitude for this process is \cite{B9}:
\begin{equation}
\emph{\textbf{A}}_{\varphi+\varphi^{*}\rightarrow \,\gamma}=-e\int d^{4}x\sqrt{-g}(f_{\vec{p}}\,(x)\stackrel{\leftrightarrow}{\partial}_{i}f_{\vec{p}\,\,'}(x))w^{*i}_{\vec{q},\lambda}(x),\label{a8}
\end{equation}
where $\sqrt{-g}=e^{3\omega t}$ and $f\stackrel{\leftrightarrow}{\partial}g = f(\partial g)-g(\partial f)$ is the bilateral derivative.
After we replace the solutions of the free scalar field from equation $(\ref{a1})$ and Maxwell field from equation $(\ref{a2})$ in the expression $(\ref{a8})$, the transition amplitude will have the following form:
\begin{eqnarray}
\emph{\textbf{A}}_{\varphi+\varphi^{*}\rightarrow \,\gamma}&=& -\frac{ie\,e^{-\pi k}}{16\sqrt{\pi}}\frac{(\vec{p}\,'-\vec{p}\,)\,\vec{\varepsilon}\,^{*}_{\lambda}(\vec{q}\,)}{\sqrt{q}\,(2\pi)^3}\int_{-\infty}^{+\infty} dt \frac{e^{-2\omega t}}{\omega}\,e^{iqt_{c}} H^{(1)}_{ik}\left(\frac{p}{\omega}e^{-\omega t}\right)H^{(1)}_{ik}\left(\frac{p\,'}{\omega}e^{-\omega t}\right)\nonumber\\
&&\times\int_{-\infty}^{+\infty}d^{3}x \, e^{i(\vec{p}+\vec{p}\,'-\vec{q}\,)\vec{x}}.
\end{eqnarray}
The spatial integral gives a delta Dirac function of the form: $(2\pi)^{3}\delta^{3}(\vec{p}+\vec{p}\,'-\vec{q}\,)$, which assures that the momentum conservation law is preserved in this process. The temporal integral contains the contribution of the expansion of the space in this process, and is expressed in terms of Hankel functions.
If we consider the new variable $z= -t_{c}$, this type of integral can be solved using equation $(\ref{a3})$ from Appendix, which relate the Hankel functions and Bessel J functions. The exponential function $e^{-iqz}$ can be expressed in terms of the modified Bessel function $K_{\frac{1}{2}}(z)$, using the relation $(\ref{a4})$ from Appendix. Then the expression of the temporal integral becomes:
\begin{eqnarray}
\frac{\sqrt{2iq}}{\sqrt{\pi}\sinh^2(\pi k)}\int_{0}^{\infty}dz \, z^{3/2} K_{\frac{1}{2}}(iqz)[J_{-ik}(pz)J_{-ik}(p\,'z)+e^{2\pi k}J_{ik}(pz)J_{ik}(p\,'z)\nonumber\\
-e^{\pi k}J_{-ik}(pz)J_{ik}(p\,'z)-e^{\pi k}J_{ik}(pz)J_{-ik}(p\,'z)].
\end{eqnarray}
The result of the integrals which contain Bessel J functions with equal indices depend on the derivative of the Legendre functions of the second kind $Q_{\nu}(u)$, with respect to the variable $u=\frac{p^2+p\,'^{2}-q^{2}}{2pp\,'}$, as one can see from the equation $(\ref{a5})$. Further using equation $(\ref{a6})$, the final result of these types of integrals can be expressed with hypergeometric Gauss function  $_{2}F_{1}$. For solving the integrals which contain Bessel J functions with opposite indices  we use the relation $(\ref{a7})$. The result for these types of integrals  depends on the Appell function $F_{4}$. The final expression for the transition amplitude is:
\begin{eqnarray}
\emph{\textbf{A}}_{\varphi+\varphi^{*}\rightarrow \,\gamma}&=& -\frac{ie(\vec{p}\,'-\vec{p}\,)}{16\sqrt{\pi}}\,\delta^{3}(\vec{p}+\vec{p}\,'-\vec{q}\,)\,\vec{\varepsilon}\,^{*}_{\lambda}(\vec{q}\,)\nonumber\\
&&\times[\textmd{g}_{k}(p,p\,',q)+ \textmd{g}_{-k}(p,p\,',q)+ \textmd{h}_{k}(p,p\,',q)+ \textmd{h}_{-k}(p,p\,',q)],\label{b1}
\end{eqnarray}
where the functions $\textmd{g}_{\pm k}(p,p\,',q)$ and $\textmd{h}_{\pm k}(p,p\,',q)$ are defined as follows:
\begin{eqnarray}
\textmd{g}_{\pm k}(p,p\,',q)&=& -\frac{i\sqrt{q}}{4(pp\,')^{3/2}}\frac{\left(k^{2}+\frac{1}{4}\right)e^{\pm\pi k}}{\cosh(\pi k)\sinh^{2}(\pi k)}\left[i e^{\pm\pi k}\,_{2}F_{1}\left(\frac{3}{2}\pm ik,\frac{3}{2}\mp ik;2;\frac{1-u}{2}\right)\right.\nonumber\\
&&\left.+\, _{2}F_{1}\left(\frac{3}{2}\pm ik,\frac{3}{2}\mp ik;2;\frac{1+u}{2}\right)\right].
\end{eqnarray}
\begin{eqnarray}
\textmd{h}_{\pm k}(p,p\,',q)= \frac{q^{-5/2}}{\pi k\sinh(\pi k)}\left(\frac{p}{p\,'}\right)^{\pm ik}F_{4}\left(\frac{3}{2}, 1, 1\pm ik, 1\mp ik; \frac{p^{2}}{q^{2}}-i0, \frac{p\,'^{2}}{q^{2}}-i0\right).
\end{eqnarray}
The probability of transition is obtained summing after the photon polarizations and squaring the transition amplitude $(\ref{b1})$. If we use the identity $|\delta^{3}(\vec{p}+\vec{p}\,'-\vec{q}\,')|^2=V\delta^{3}(\vec{p}+\vec{p}\,'-\vec{q}\,')$, the final result of the probability density is:
\begin{eqnarray}
\emph{\textbf{P}}_{\varphi+\varphi^{*}\rightarrow \,\gamma}&=& \frac{1}{2}\sum_{\lambda}|\emph{\textbf{A}}_{\varphi+\varphi^{*}\rightarrow \,\gamma}|^{2}= \frac{1}{2}\sum_{\lambda}\frac{e^{2}\,V}{256 \pi }|(\vec{p}\,'-\vec{p}\,)\vec{\varepsilon}\,^{*}_{\lambda}(\vec{q}\,)|^{2}\delta^{3}
(\vec{p}+\vec{p}\,'-\vec{q}\,)\nonumber\\
&&\left[|\textmd{g}_{k}(p,p\,',q)|^{2}+ |\textmd{g}_{-k}(p,p\,',q)|^{2}+ |\textmd{h}_{k}(p,p\,',q)|^{2}+ |\textmd{h}_{-k}(p,p\,',q)|^{2} \right.\nonumber\\
&& +\left.\textmd{g}^{*}_{k}(p,p\,',q)\textmd{g}_{-k}(p,p\,',q)+ \textmd{g}^{*}_{k}(p,p\,',q)\textmd{h}_{k}(p,p\,',q)+ \textmd{g}^{*}_{k}(p,p\,',q)\textmd{h}_{-k}(p,p\,',q)\right.\nonumber\\
&&+\left.\textmd{g}^{*}_{-k}(p,p\,',q)\textmd{g}_{k}(p,p\,',q)+ \textmd{g}^{*}_{-k}(p,p\,',q)\textmd{h}_{k}(p,p\,',q)+ \textmd{g}^{*}_{-k}(p,p\,',q)\textmd{h}_{-k}(p,p\,',q)\right.\nonumber\\
&&+\left.\textmd{h}^{*}_{k}(p,p\,',q)\textmd{g}_{k}(p,p\,',q)+ \textmd{h}^{*}_{k}(p,p\,',q)\textmd{g}_{-k}(p,p\,',q)+ \textmd{h}^{*}_{k}(p,p\,',q)\textmd{h}_{-k}(p,p\,',q)\right.\nonumber\\
&&+ \left.\textmd{h}^{*}_{-k}(p,p\,',q)\textmd{g}_{k}(p,p\,',q)+ \textmd{h}^{*}_{-k}(p,p\,',q)\textmd{g}_{-k}(p,p\,',q)+ \textmd{h}^{*}_{-k}(p,p\,',q)\textmd{h}_{k}(p,p\,',q)\right].  \label{q3} \nonumber\\
\end{eqnarray}
We observe that the transition probability is proportional with a term of the form: \,\,\,\,\,\, $\sum_{\lambda}|(\vec{p}\,'-\vec{p}\,)\vec{\varepsilon}\,^{*}_{\lambda}(\vec{q}\,)|^{2}$. Using the momentum conservation $\vec{p}\,\,'=\vec{q}-\vec{p}$ and equation $(\ref{b2})$, this term can be expressed as follows \cite{B9}:
\begin{equation}
\sum_{\lambda}|(\vec{p}\,'-\vec{p}\,)\vec{\varepsilon}\,^{*}_{\lambda}(\vec{q}\,)|^{2}=4p^{2}\sin^2{\theta_{qp}}=\frac{4p\,'^{2}p^{2}}{q^{2}}\sin^{2}{\theta_{pp\,'}
}.
\end{equation}
The final result of the probability of transition depends on the parameter $x=m/\omega$ and the particle momenta $p\,,p\,',q$. This allows us to study the behaviour of the probability in terms of $m/\omega$ for fixed momenta. It is also worth to mention that, using the perturbative approach the probability density  is a quantity that depends on the momenta of the particles. Consequently, a divergence in the integrals over the final momenta could be solved using regularization methods.

\section{The total probability}
The total probability is obtained integrating the transition probability $(\ref{q3})$, after the final momenta of the photon:
\begin{equation}
\emph{\textbf{P}}_{tot}=\int d^{3}q\, \emph{\textbf{P}}_{\varphi+\varphi^{*}\rightarrow \,\gamma}.
\end{equation}
In the case of scalar pair annihilation, the expression of the total probability can be evaluated analytically, because we have to perform only one integral over the final momentum of the photon.
Taking into account that the result of the probability of transition is proportional with delta-Dirac function $\delta^{3}(\vec{p}+\vec{p}\,'-\vec{q}\,)$, the momentum integration can be easily performed. Then the final result for the total probability per unit of volume reads:
\begin{eqnarray}
\frac{\emph{\textbf{P}}_{tot}}{V}&=& \frac{e^{2}\sin^{2}\theta_{pp\,'}}{128\pi}\frac{p^{2}p\,'^2}{|\vec{p}+\vec{p}\,'|^2}
\left[|\textmd{g}_{k}(p,p\,')|^{2}+ |\textmd{g}_{-k}(p,p\,')|^{2}+ |\textmd{h}_{k}(p,p\,')|^{2}+ |\textmd{h}_{-k}(p,p\,')|^{2} \right.\nonumber\\
&& +\left.\textmd{g}^{*}_{k}(p,p\,')\textmd{g}_{-k}(p,p\,')+ \textmd{g}^{*}_{k}(p,p\,')\textmd{h}_{k}(p,p\,')+ \textmd{g}^{*}_{k}(p,p\,')\textmd{h}_{-k}(p,p\,')\right.\nonumber\\
&&+\left.\textmd{g}^{*}_{-k}(p,p\,')\textmd{g}_{k}(p,p\,')+ \textmd{g}^{*}_{-k}(p,p\,')\textmd{h}_{k}(p,p\,')+ \textmd{g}^{*}_{-k}(p,p\,')\textmd{h}_{-k}(p,p\,')\right.\nonumber\\
&&+\left.\textmd{h}^{*}_{k}(p,p\,')\textmd{g}_{k}(p,p\,')+ \textmd{h}^{*}_{k}(p,p\,')\textmd{h}_{-k}(p,p\,')+ \textmd{h}^{*}_{k}(p,p\,')\textmd{g}_{-k}(p,p\,')\right.\nonumber\\
&&+ \left.\textmd{h}^{*}_{-k}(p,p\,')\textmd{g}_{k}(p,p\,')+ \textmd{h}^{*}_{-k}(p,p\,')\textmd{g}_{-k}(p,p\,')+ \textmd{h}^{*}_{-k}(p,p\,')\textmd{h}_{k}(p,p\,')\right]. \,\,\,\,\,\,\,\,\,\,\,\, \label{nr2}
\end{eqnarray}
The expressions of $\textmd{g}_{\pm k}(p,p\,')$ and $\textmd{h}_{\pm k}(p,p\,')$ which are now functions only of $p$ and $p\,'$ are:
\begin{eqnarray}
\textmd{g}_{\pm k}(p,p\,')&=& -\frac{i|\vec{p}+\vec{p}\,'|^{1/2}}{4(pp\,')^{3/2}}\frac{\left(k^{2}+\frac{1}{4}\right)e^{\pm\pi k}}{\cosh(\pi k)\sinh^{2}(\pi k)}\left[i e^{\pm\pi k}\,_{2}F_{1}\left(\frac{3}{2}\pm ik,\frac{3}{2}\mp ik;2;\frac{1+\cos{\theta_{pp\,'}}}{2}\right)\right.\nonumber\\
&&\left.+\, _{2}F_{1}\left(\frac{3}{2}\pm ik,\frac{3}{2}\mp ik;2;\frac{1-\cos{\theta_{pp\,'}}}{2}\right)\right]\label{nr1}
\end{eqnarray}

\begin{eqnarray}
\textmd{h}_{\pm k}(p,p\,')=\frac{|\vec{p}+\vec{p}\,'|^{-5/2}}{\pi k\sinh(\pi k)}\,\left(\frac{p}{p\,'}\right)^{\pm ik}\,F_{4}\left(\frac{3}{2}, 1, 1\pm ik, 1\mp ik; \frac{p^{2}}{|\vec{p}+\vec{p}\,'|^{2}}-i0, \frac{p\,'^{2}}{|\vec{p}+\vec{p}\,'|^{2}}-i0\right),\label{nr2}\nonumber\\
\end{eqnarray}
where $|\vec{p}+\vec{p}\,'|=(p^2+p\,'^2+2pp\,'\cos{\theta_{pp\,'}})^{1/2}$.

Further we want to graphically analyse the total probability of photon emission as a function of $x=m/\omega\in (1.5,\infty)$, for different values of the angle $\theta_{pp\,'}$, keeping $p$ and $p\,'$ fixed. As can be seen from $(\ref{nr1})$ the hypergeometric function $_{2}F_{1}$ is divergent for $\theta_{pp\,'}=0,\pi$. In this situation we will study the behaviour of the total probability for small angles $(\pi/25\leq\theta_{pp\,'}\leq\pi/15)$ as well as for large angles $(0.9\pi\geq\theta_{pp\,'}\geq 0.72\pi)$.  The $h_{\pm k}$ functions are proportional with a factor $(k\sinh(\pi k))^{-1}$, which make them convergent for large $m/\omega$. The relevant contribution at large $m/\omega$ will be given by this factor and for that reason we approximate the Appell functions $F_{4}\simeq1$.

\begin{figure}[h!t]
\centerline{\includegraphics[scale=0.4]{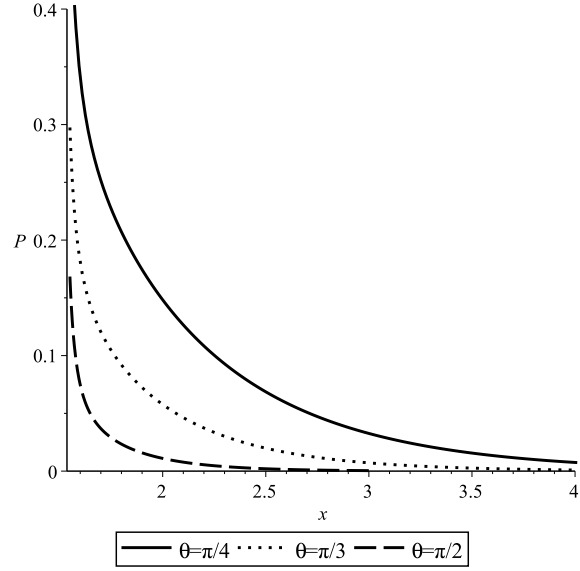}}
\caption{The total probability $P$ as function of $x=m/\omega$, for $p=1, p\,'= 2$ and $\theta_{pp\,'}\in\{\frac{\pi}{4},\frac{\pi}{3},\frac{\pi}{2}\}.$}
\label{fig:1}
\end{figure}
\newpage

\begin{figure}[h!t]
\centerline{\includegraphics[scale=0.4]{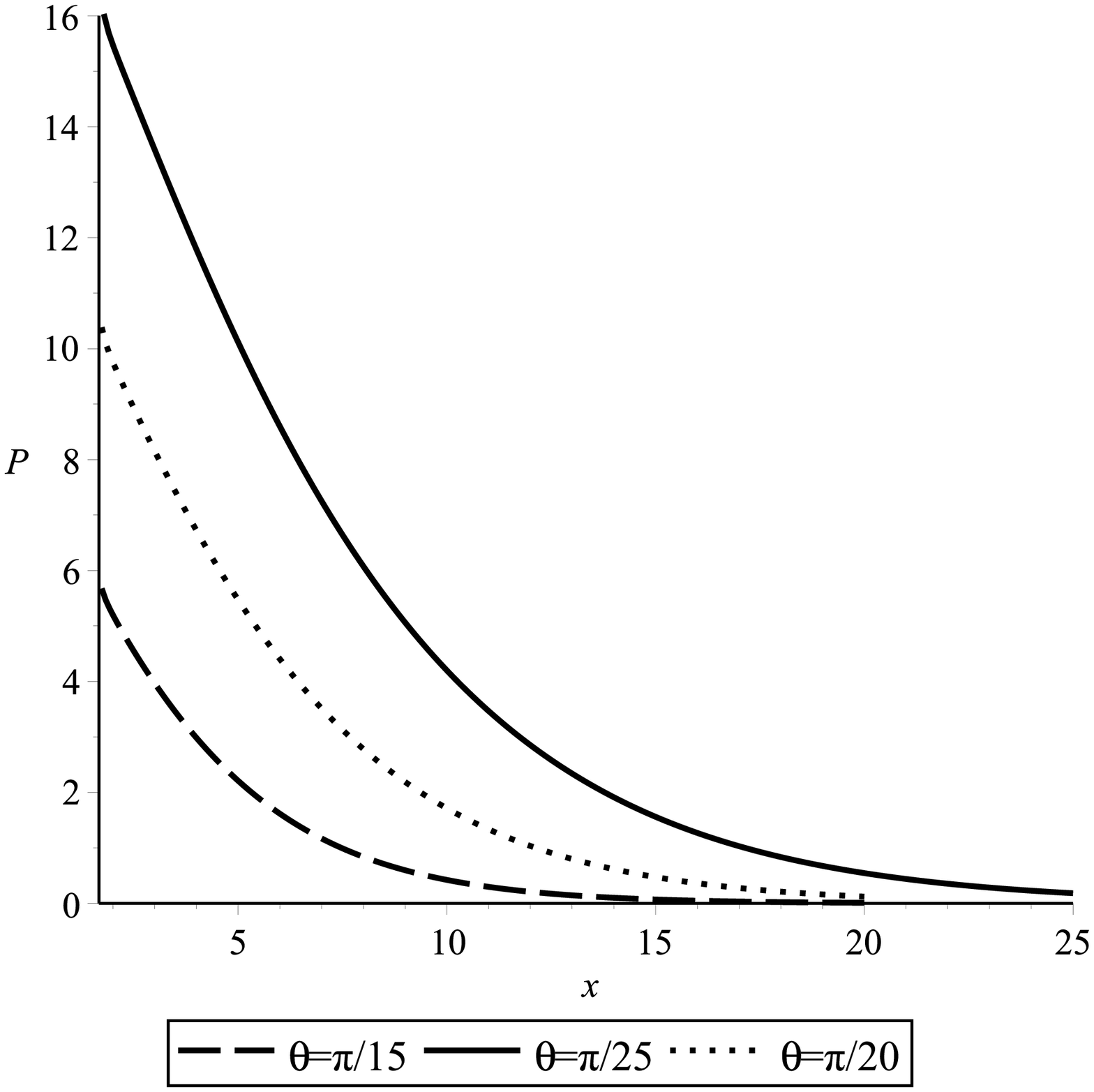}}
\caption{The total probability $P$ as a function of $x=m/\omega$, for $p=1, p\,'=2$ and small values of the angle $\theta_{pp\,'}$.}
\label{fig:2}
\end{figure}

\begin{figure}[h!t]
\centerline{\includegraphics[scale=0.4]{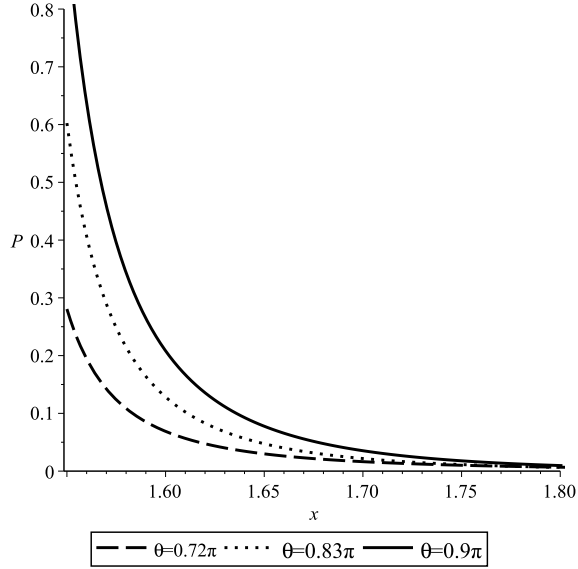}}
\caption{The total probability $P$ as a function of $x=m/\omega$, for $p=1, p\,'= 2$ and large values of the angle $\theta_{pp\,'}$.}
\label{fig:3}
\end{figure}

\newpage
From Fig.1 it can be seen that, in the particular cases when $\theta_{pp\,'}\in\{\frac{\pi}{4},\frac{\pi}{3},\frac{\pi}{2}\}$ the total probability is a function which decreases as the parameter $m/\omega$ increases. This property of the total probability is also preserved for  the small/large angle cases. The obvious conclusion is that the total probability of  scalar pair annihilation is significative only for $m/\omega\in(1.5,2.5)$ where the gravitational fields are still strong. From Fig.2 we see that, in the case when $\theta_{pp\,'}$ is close to zero, the total probability is large compared to the other studied cases. This important observation must be interpreted by analysing the analytical formula $(\ref{nr2})$ of the total probability. As it can be seen from $(\ref{nr2})$ the total probability of scalar pair annihilation is proportional with $\sin^{2}{\theta_{pp\,'}}$, which arises from the polarization term. One is tempted to say that when $\theta_{pp\,'}\rightarrow 0$, the total probability also approaches zero, but the functions $g_{\pm k}(p,p\,')$ have a divergent behaviour for $\theta_{pp\,'}=0$, because the Gauss hypergeometric function $_{2}F_{1}$ becomes also divergent in this case. Also in the large angle case, the total probability of massive scalar pair annihilation increases when $\theta_{pp\,'}\rightarrow \pi$, but this probability is small comparatively with the small angle case. This result shows that, it is more probable for the scalar pair to annihilate into a photon when the momenta of the scalar particles are parallel and  orientated in the same direction. Also from these graphs we can draw the conclusion that the annihilation probability is minimal when $\theta_{pp\,'}=\frac{\pi}{2}$.
Another important observation which can be made from  Figs. $(1)-(3)$ is that, the total probability of scalar pair annihilation vanishes in the Minkowski limit, when the parameter $m/\omega\rightarrow\infty$. The Minkowski limit can be also verified by analytical calculations. From the expressions $(\ref{nr1})-(\ref{nr2})$ it can be seen that when $m/\omega\gg 1$, the functions $g_{k}(p,p\,')$ and $h_{k}(p,p\,')$ decreases rapidly to zero with a term of the form $e^{-\pi\frac{m}{\omega}}$.

It seems from our study that the total probability of the created photons is divergent for $\theta_{pp\,'}=0,\pi$. A possible way to eliminate the divergence is by using regularization methods, since the density of probability depends on the final momenta.

\section{Conclusion}

In this letter we have studied the annihilation of massive scalar particles into a photon in de Sitter spacetime, using a direct perturbative calculation. We were able to obtain an analytical formula for the total probability of scalar pair annihilation. The total probability of created photons is a function which depends on the parameter $m/\omega$ and the momenta of the scalar particles. Also, the graphical results show that the total probability is large when the scalar particles annihilate at small angles. In the limit $\omega\rightarrow\,0$  we recover the flat space result where the probability for this process vanish.

\section{Appendix}

For solving our temporal integral we express the Hankel functions in terms of Bessel J functions as follows \cite{B20}:
\begin{eqnarray}
H^{(1)}_{\mu}(z)&=&\frac{J_{-\mu}(z)-e^{-i\pi\mu}J_{\mu}(z)}{i\sin{\pi\mu}}\nonumber\\
H^{(2)}_{\mu}(z)&=&\frac{e^{i\pi\mu}J_{\mu}(z)-J_{-\mu}(z)}{i\sin{\pi\mu}}.\label{a3}
\end{eqnarray}
To replace the exponential from our integrals, we use the expression of Bessel K function of index $\frac{1}{2}$ \cite{B20}:
\begin{equation}
K_{\frac{1}{2}}(z)=\sqrt{\frac{\pi}{2z}}e^{-z},\label{a4}
\end{equation}
The integrals which contain products of Bessel J functions with equal indices can be expressed in terms of the Legendre functions of second kind $Q_{\nu}(u)$, where $2abu= a^{2}+b^2+c^2$ \cite{B20}:
\begin{eqnarray}
\int_{0}^{\infty}dx x^{\frac{3}{2}}K_{\frac{1}{2}}(cx)J_{\nu}(ax)J_{\nu}(bx)= -\frac{1}{\sqrt{2\pi}}\frac{c^{\frac{1}{2}}}{(ab)^{3/2}}\frac{d}{du}Q_{\nu-\frac{1}{2}}(u)\label{a5}
\end{eqnarray}
The integrals which contain products of two Bessel J functions with opposite indices can be solved using the relation \cite{B20}:
\begin{eqnarray}
\int_{0}^{\infty}dx x^{\frac{3}{2}}K_{\frac{1}{2}}(cx)J_{\nu}(ax)J_{-\nu}(bx)= \frac{\sin{\pi\nu}}{\sqrt{2\pi}\nu}c^{-\frac{5}{2}}\left(\frac{a}{b}\right)^{\nu}F_{4}\left(\frac{3}{2},1,1+\nu,1-\nu;-\frac{a^{2}}{c^{2}},-\frac{b^{2}}{c^{2}}\right),\label{a7}
\end{eqnarray}
where $F_{4}$ is the Appell hypergeometric function with double argument.
The above integrals are convergent for $\mathcal{R}e(c)>0$ and for solving the integrals we take $c\rightarrow\epsilon-iq$ with $\epsilon>0$ and finally we consider the limit $\epsilon\rightarrow\,0$.

The derivative of the Legendre function $Q_{\nu}(u)$ in respect to $u$ can be written in terms of the Gauss hypergeometric function $_{2}F_{1}$ as follows \cite{B7}:
\begin{eqnarray}
\frac{d}{du}Q_{\nu}(u\pm i0)= \frac{\pi\nu(\nu+1)}{4\sin{\pi\nu}}\left[e^{\mp i\pi\nu}\,_{2}F_{1}\left(1-\nu,2+\nu;2;\frac{1-u}{2}\right)+\,_{2}F_{1}\left(1-\nu,2+\nu;2;\frac{1+u}{2}\right)\right].\label{a6}
\end{eqnarray}
This relation help us to establish the final form of the transition amplitude.

\par
\textbf{Acknowledgements}
\par
This work was supported by the strategic grant POSDRU/159/1.5/S/137750, Project "Doctoral and Postdoctoral programs support for increased competitiveness in Exact Sciences research" cofinanced by European Social Fund within the Sectoral Operational Programme Human Resources Development 2007-2013.

\end{document}